\documentclass[apjl]{emulateapj}

\begin{document}

\author{Amit K. Misra\altaffilmark{1,2,3}}
\author{Victoria S. Meadows\altaffilmark{1,2,3}}

\altaffiltext{1}{Astronomy Department, University of Washington, Seattle, 98195}
\altaffiltext{2}{University of Washington Astrobiology Program}
\altaffiltext{3}{NASA Astrobiology Institute Virtual Planetary Laboratory}

\title{Discriminating Between Cloudy, Hazy and Clearsky Exoplanets Using Refraction}

\begin{abstract}

We propose a method to distinguish between cloudy, hazy and clearsky (free of clouds and hazes) exoplanet atmospheres that could be applicable to upcoming large aperture space and ground-based telescopes such as the James Webb Space Telescope (JWST) and the European Extremely Large Telescope (E-ELT). These facilities will be powerful tools for characterizing transiting exoplanets, but only after a considerable amount of telescope time is devoted to a single planet. A technique that could provide a relatively rapid means of identifying haze-free targets (which may be more valuable targets for characterization) could potentially increase the science return for these telescopes. Our proposed method utilizes broadband observations of refracted light in the out-of-transit spectrum. Light refracted through an exoplanet atmosphere can lead to an increase of flux prior to ingress and subsequent to egress. Because this light is transmitted at pressures greater than those for typical cloud and haze layers, the detection of refracted light could indicate a cloud- or haze-free atmosphere. A detection of refracted light could be accomplished in $<$10 hours for Jovian exoplanets with JWST and $<$5 hours for Super-Earths/Mini-Neptunes with E-ELT. We find that this technique is most effective for planets with equilibrium temperatures between 200 and 500 K, which may include potentially habitable planets. A detection of refracted light for a potentially habitable planet would strongly suggest the planet was free of a global cloud or haze layer, and therefore a promising candidate for follow-up observations.

% If we have the words available after all these edits, can you add a sentence or phrase expanding on the results?   e.g.  "We find that this effect is likely to be most pronounced for planets with equilibrium temperatures less than 300K, which would include potentially habitable planets".  Detection of this refracted light etc....  Or something like that?   I know that potentially habitable planets, being terrestrials, probably have lousy detectability, even if they are great at refracting light, so maybe that is a misleading statement, but it would be nice to mention the distance effect and the fact that you've found a transit transmission behavior that actually IMPROVES with distance from the star. 

%Light is refracted through an exoplanet atmosphere and this can lead to an increase of flux prior to ingress and subsequent to egress.

%We find that detecting refracted light, and therefore characterizing an exoplanet as haze-free, could be accomplished in $<$10 hours for Jovian exoplanets with JWST and $<$5 hours for Super-Earths/Mini-Neptunes with E-ELT.    .... I chopped this down to the bare "final credibility statement" because the last two sentences were repeating each other quite a bit.  

%Didn't like "clearsky exoplanet" because it's a bit jargon-ey, if understandable

\end{abstract}


\section{Introduction} 

Transit transmission spectroscopy is an observational technique that can be used to characterize a planet's atmosphere as it transits its host star. This technique has been used to identify absorption features in some exoplanet atmospheres \citep{char02,vidal03, barman07, desert08, deming13}, but many planets have flat, featureless spectra \citep{kreidberg14, knutson14}. Flat spectra can be explained by either a high mean molecular weight (and thus small scale height) atmosphere or by the presence of high-altitude clouds or hazes, as is common in planets in our own Solar System, and has been inferred for the atmospheres of some Hot Jupiters \citep{sing11, sing13}. For GJ 1214b, even high mean molecular weight atmospheres have recently been ruled out, leaving very high altitude clouds or hazes as the only physically plausible explanation for the planet's spectrum \citep{kreidberg14}.

%previous version of 2nd paragraph
%The James Webb Space Telescope (JWST) and large ground based telescopes such as the European Extremely Large Telescope (E-ELT) will open up new avenues for characterizing transiting planets. JWST may be able to detect absorption features for an Earth-like or super-Earth planet in the near future by using $\sim$200 hr of in-transit observations \citep{deming09, misra14a}. Detecting features such as H$_2$O and CO$_2$ could be possible for an Earth-like planet orbiting an F, G or K star with E-ELT in $\sim$10 transits \citep{hedelt13}. It may be possible to detect the O$_2$ A band with the E-ELT with 20 hours of in-transit time over the span of 2 years or more \citep{rodler14}. If there are clouds or hazes present in an atmosphere, it will limit the atmospheric levels that can be probed, making the targets less desirable for characterization. Therefore, it would be beneficial to have a method that could relatively rapidly discriminate between cloud and haze-free planets, and cloudy or hazy planets, which are not easily characterized even in extended transit transmission observations \citep{kreidberg14}. 

The James Webb Space Telescope (JWST) and large ground based telescopes such as the European Extremely Large Telescope (E-ELT) will open up new avenues for characterizing transiting planets. Absorption features for an Earth-like or super-Earth planet could be detected in the near future with 200 hrs of JWST in-transit observations \citep{deming09, misra14a}, or with $>$20 hrs of E-ELT in-transit observations \citep{hedelt13, rodler14}. If there are clouds or hazes present in an atmosphere, they will limit the atmospheric levels that can be probed, making the targets less desirable for characterization. Therefore, it would be beneficial to have a method that could relatively rapidly discriminate between haze-free and hazy planets, which are not easily characterized even in extended transit transmission observations \citep{kreidberg14}. 

Here we examine whether refractive effects on transit transmission spectroscopy could provide a more efficient way of discriminating between hazy, cloudy and clearsky (free of clouds and hazes) atmospheres. \citet{benneke13} propose that measurements of absorption wing steepness, or a comparison of the depths of multiple absorption bands, could be used to distinguish between cloudy/hazy and clearsky planets. Since both methods require relatively detailed characterization of absorption features, these techniques may not discriminate between a hazy and haze-free planet before considerable amounts of telescope time are used. In contrast, the refractive signal is independent of absorption features, and could be binned over a wide range of wavelengths, increasing detectability. While refraction can set a mid-transit maximum transit pressure (or minimum tangent altitude) that can be probed by transit transmission spectroscopy \citep{garcia12, betremieux14, misra14b} refraction provides the deepest probe of an atmosphere pre- and post-transit \citep{misra14b}, when it also generates a refractive halo around the exoplanet, increasing the observed flux \citep{sidis10, garcia12, garcia12b}. \citet{sidis10} derive analytic expressions for the halo brightness for both transparent atmospheres and atmospheres with extinction from Rayleigh scattering. \citet{garcia12} and \citet{garcia12b} examine the concept further by generating spectra of refracted light for the Earth and for Venus.

Here we expand on previous work by showing that a detection of refracted light in a transit light curve pre-ingress and post-egress would preclude hazy atmospheres, because hazes tend to obscure the layers of the atmosphere that refract light to a distant observer. We show that this signal could be more readily detectable than spectral absorption features in some cases and could be valuable for selecting targets for more extended follow-up observations.

\section{Methods}

\subsection{Model Description} \label{sec:modeldesc}

We used the refraction code that is described in detail in \citet{misra14b} to calculate refraction angles for a suite of planetary atmospheres. Briefly, refraction is governed by a set of differential equations that we solve at each step along the path through the atmosphere using a Runge-Kutta integration scheme. Given the planetary radius, surface gravity, atmospheric composition and pressure-temperature profile, the model calculates the angle of deflection due to refraction for a range of tangent altitudes.

We generated refractive light curves to calculate the amount of out-of-transit refracted light. We first determined whether or not each portion of the atmosphere (given as an altitude and angle along the annulus of the atmosphere) is illuminated at each time during the transit event, from half a transit length prior to ingress to half a transit length after egress, and then integrated over the entire atmosphere to generate the light curve.

\begin{figure}
\includegraphics[width=8cm]{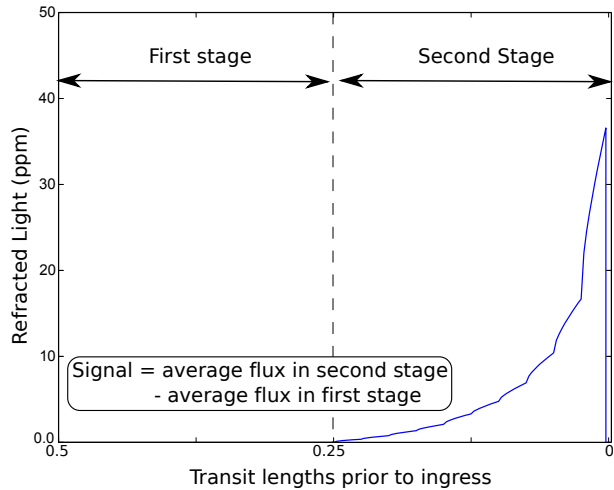}
\caption{Pre-transit light curve for a 300 K Saturn-analog orbiting a Sun-like star from half a transit length prior to ingress to ingress. Refraction leads to an increase in flux prior to ingress (and subsequent to egress by symmetry). This particular case gives the greatest peak brightness for all cases orbiting Sun-like stars. We define the signal of refracted light as the average flux level in stage 2 (just prior to ingress) minus the average flux level in stage 1.}
\label{fig:lightcurve}
\end{figure}

We quantified the signal of refracted light as the difference in the average value of the transit light curve between two stages of the transit event. We chose a quarter of a transit length as the time bin to maximize the signal to noise ratio (S/N) for the majority of cases we examined. As can be seen in Figure \ref{fig:lightcurve}, most of the refracted flux is seen in the quarter of a transit prior to ingress, so dividing the transit into longer stages would reduce the time-averaged signal. Stages with shorter durations could increase the time-averaged signal, but would have greater noise levels because of the shorter integration time. Refracted light brightness is more strongly peaked just outside of transit for planets with equilibrium temperatures (T$_{eq}$, see \citet{borucki11} for definition) $>$600 K, but we find that even for these cases adopting a time bin of 5\% of the transit length results in poorer S/N for T$_{eq}<$600 K and an increase in S/N by only a factor of $\sim$2 for planets with greater temperatures

%shorter times -> shorter durations

%As shown in Figure \ref{fig:lightcurve}, the first stage is from half a transit to a quarter of a transit length prior to ingress (or subsequent to egress due to symmetry), and the second stage is from a quarter of a transit length prior to ingress until ingress. Because refraction leads to greater flux values just prior to ingress, the flux levels in the second stage will be greater than those of the first stage, and that difference is the signal we used in determining the detectability of refracted light. 

\subsection{Test Cases} \label{sec:cases}

We used a suite of planetary atmospheres to calculate the refracted light signal. These are shown in Table \ref{tab:params}. We have selected a combination of solar system analogs as well possible super-Earth and mini-Neptune atmospheres to cover a wide range of potential planetary atmospheres. We assumed the H$_2$-dominated atmospheres have a solar H/He ratio (90\% H, 10\% He) for simplicity, but the small change in the refractive index for different H/He ratios should have a negligible effect on our results. For the super-Earth and mini-Neptune planets, we ran our models on 4 test cases to span the most likely bulk atmospheric compositions: 100\% N$_2$, solar composition, 100\% H$_2$O, and 100\% CO$_2$.

Out-of-transit refracted light must be deflected by a large enough angle to be scattered into the beam to a distant observer. The characteristic angle of deflection (in radians) is $\sim$R$_*$/d, (where R$_*$ is the stellar radius and $d$ is the planet-star distance) which is also half the angular size of the star, as seen by the planet. For example, half a transit length prior to ingress, on the trailing side of the planet, light originating at the near and far limb of the star would have to be refracted by R$_*$/d and 3R$_*$/d respectively, to reach a distant observer. More than half a transit length prior to ingress, the required refraction angles would increase, and closer to ingress they would decrease.

% added planet-star distance definition above since this is the first time you introduce it. 

% as suggested by the reviewer, we need a line in here about this being very sensitive to "d" and so often overlooked for hot Jupiters ...but perhaps you have that in the discussion later?   Yes you do!   So ignore this comment.  

Based on the qualitative description given above, the brightness of the refracted light signal depends on the angles of refraction at each altitude in an atmosphere and the planet-star geometry. The deflection of light by a planetary atmosphere can be calculated by our model from the atmospheric scale height, the planetary radius (R$_p$), and the index of refraction of the atmosphere. For each test case, R$_p$ and the refractive index are given. The scale height is determined from the surface gravity, mean molecular weight of the atmosphere, and T$_{eq}$. Surface gravity is given for each test case, and the mean molecular weight is determined by the composition. We ran our model simulations over a grid of isothermal atmospheres with T$_{eq}$ from 100 to 1000 K, covering a wide range of atmospheric scale heights. We chose to use isothermal atmospheres for simplicity after testing other temperature profiles with realistic tropospheric lapse rates and stratospheric temperature inversions and finding no significant difference in our results. The planet-star geometry is determined by R$_p$, R$_*$, the impact parameter ($b$), and $d$. The impact parameter is the sky-projected distance at conjunction, in units of stellar radius \citep{winn11}. To cover the full range of planet-star geometries, we ran our simulations over a range of values for $b$, planetary albedo, and stellar types from M9 to F5, constraining the stellar radius and luminosity.

%why do you state that the first two quantities are given above, since you describe that mean molecular weight is calculated, not given? 

\begin{deluxetable*}{lcccc}
\tablecaption{Planetary Atmosphere Test Cases}
\tablehead{ \colhead{Planet Type} & \colhead{Radius (km)} & \colhead{Composition} & \colhead{Refractive Index at STP} & {Surface Gravity (m s$^{-2}$)} }
\startdata
Earth & 6371 & N$_2$ & 1.00029 & 9.8 \\
Super-Earth & 12742 & N$_2$ & 1.00029 & 9.8 \\
Mini-Neptune & 12742 & H$_2$ & 1.00012 & 9.8\\
H$_2$O Super-Earth & 12742 & H$_2$O & 1.00026 & 9.8\\
CO$_2$ Super-Earth & 12742 & CO$_2$ & 1.00044 & 9.8 \\
Neptune & 24622 & H$_2$ & 1.00012 & 11.1 \\
Saturn & 58232 & H$_2$ & 1.00012 & 10.44\\
Jupiter & 69911 & H$_2$ & 1.00012 & 24.8\\
\enddata
\label{tab:params}
\end{deluxetable*}

Because our model does not explicitly calculate the effect of cloud and aerosol opacity, we simulated the effect of a cloud or haze layer by truncating the depth of the measurable atmosphere at a characteristic pressure layer. To determine appropriate pressure cutoff layers for the three main aerosol cases under consideration, we used our modeling results and examples of clouds and hazes in our own Solar System to select pressure cutoffs at 1 bar (clearsky case), 0.1 bars (cloudy case) and 1 mbar (hazy case). We chose 1 bar as our clearsky pressure cutoff because at pressures $\geq$1 bar, our modeling indicates that atmospheres within the range of compositions under consideration are optically thick near 1 $\mu$m (the central wavelength for our transit simulations) when only Rayleigh scattering is included. For the pressure cut-off for cloudy atmospheres we chose 0.1 bars, which is a characteristic lower pressure limit for the tropopause for atmospheres of a range of different compositions \citep{robinson14}, and the majority of clouds are found within a planet's troposphere. Lastly, we chose 1 mbar as the hazy pressure cutoff because hazes are typically generated via photochemistry in the upper atmosphere at pressures near 1 mbar. For example, at 1 $\mu$m Venus is optically thick ($\tau$=1) in transit transmission at 90 km ($\sim$0.1 mbar) \citep{ehrenreich12} and Titan is optically thick at $\sim$240 km ($<$0.5 mbar) \citep{bellucci09}. Because hazes form at pressures $<$1 mbar in both a warm CO$_2$-dominated atmosphere and a cold N$_2$-dominated atmosphere, we chose 1 mbar as a reasonable cutoff for hazes over the parameter space we explore here.

\subsection{Detectability}

We used the publicly available exposure time calculators (ETCs) to estimate the S/N for detecting refracted light in a transit light curve with JWST\footnote{http://jwstetc.stsci.edu/} and E-ELT\footnote{https://www.eso.org/observing/etc/}. We used the ETCs to estimate the noise at 1 $\mu$m with spectral resolving power (R) equal to 100 for stellar types from F5V to M9V. We chose the lowest resolving power available (R=100 for the JWST ETC) because this technique does not require high resolving power, and could even be performed with broadband filter photometry if necessary. The input stellar spectra were Phoenix NextGen spectra with solar metallicities \citep{hauschildt99} with the star placed at a distance of 10 pc. Given the parts per million (ppm) flux difference for refracted light and the estimated S/Ns from the ETCs, we calculated the out-of-transit integration time required to detect refracted light at a S/N$>$3 over all T$_{eq}$ values for each planetary atmosphere and stellar type.

\section{Results}

\begin{deluxetable*}{lccccccccccc}
\tablecaption{Refracted Light Signals} 
\tablehead{\colhead{Planet} & \colhead{T$_{eq}$} & \colhead{T$_*$}  &  \colhead{Atm.}  & \colhead{Albedo}  &  \colhead{Flux} & \colhead{Int.}   &\colhead{E-ELT} & \colhead{Tot.} & \colhead{Int.} & \colhead{JWST} & \colhead{Tot.} \\
\colhead{Type} & \colhead{(K)} &   \colhead{(K)} & \colhead{Type}  & & {(ppm)} & {Time (h)} & {Transits} & {Time (yr)} & {Time (h)} & {Transits} & {Time (yr)}}
\startdata
Earth        &       400  & 5780 & Clearsky  & 0.15    &  0.13&  4.77  &   1.0    & 0.3  & 999.00  & 781.2  & 231.9\\
Super-Earth &   450 &  5780 & Clearsky  & 0.15     & 0.29 & 0.91    & 1.0    & 0.2  & 999.00  & 166.9   & 34.8\\
Mini-Neptune         &   250 &  5780&  Clearsky &  0.15     & 1.98 & 0.02    & 1.0   &  1.2   & 27.65    & 2.0   &  2.4\\
H$_2$O Super-Earth        &   400  & 5780 & Clearsky &  0.15   &    0.41&  0.45  &   1.0 &    0.3  & 640.78  &  73.9  &  21.9\\
CO$_2$ Super-Earth         &  600  & 5780 & Clearsky  & 0.15    &  0.22&  1.60  &   1.0   &  0.1  & 999.00  & 393.7   & 34.6\\
Neptune   &          250 &  5780&  Clearsky  & 0.15    &  3.79 &  0.01  &   1.0  &   1.2   &  7.58  &   1.0   &  1.2\\
Saturn      &        300 &  5780 & Clearsky &  0.15   & 10.98 & 0.01   &  1.0  &   0.7 &    0.90 &    1.0   &  0.7\\
Jupiter          &   350 &  5780 & Clearsky &  0.15   &   6.39 & 0.01   &  1.0    & 0.4    & 2.66   &  1.0 &    0.4\\
\enddata
\tablecomments{The E-ELT results were calculated assuming 50 spectral resolution elements could be binned over. Results shown here are for most favorable cases orbiting Sun-like stars.}
\label{tab:results}
\end{deluxetable*}

Table \ref{tab:results} shows the ppm flux change, and the required integration time, number of transits and total time (from first transit to last) for detecting refracted light for each test case over the suite of parameters. The results shown here are for an albedo of 0.15, but results for other albedos are available online. Our results indicate that Saturn analog planets exhibit the most detectable refracted light of any of the cases because Saturn has a radius close to Jupiter's radius and a lower surface gravity, which increases the atmospheric scale height at a given temperature. The amplitude of the refracted light signal (as defined in Section \ref{sec:modeldesc}) is no larger than half a scale height for all cases we have explored here. The maximum flux amplitude for planets orbiting Sun-like stars is 10 ppm for a 300 K Saturn analog. The other H$_2$ cases have maximum amplitudes of 6, 4, and 2 ppm for the Jupiter, Neptune, and Mini-Neptune cases, respectively. The greatest ppm signals are for planets orbiting around M9V stars, for which the signals can increase by nearly two orders of magnitude to 950 ppm for a 200 K Saturn analog. 

\begin{figure}[h]
\includegraphics[width=9cm]{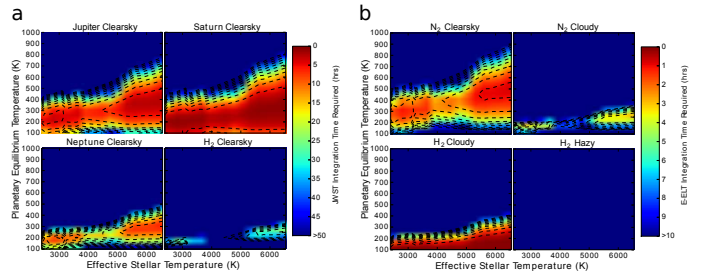}
\caption{\textbf{a)} JWST integration time required to detect refracted light for H$_2$-dominated atmospheres. The results assume that observations are made at 1 $\mu$m with R=100, for a planet at a distance of 10 pc. The planets with the most detectable refracted light signal are those with T$_{eq}<$600 K for solar-type stars and $<$400 K for M dwarfs. For many Jupiter and Saturn-analog cases, refracted light could be detected with $<$10 hour of JWST time. \textbf{b)} E-ELT integration time required to detect refracted light for N$_2$ and H$_2$ atmospheres, assuming that 50 wavelength bins can be summed over at R=100. Refracted light is most detectable for the non-hazy atmospheres, and therefore could be used to distinguish between hazy and non-hazy worlds. }
\label{fig:inttime}
\end{figure}

Figure \ref{fig:inttime}A shows the JWST out-of-transit integration time required to detect refracted light for the 4 H$_2$-dominated atmospheres: Jupiter, Saturn, Neptune analogs and the `mini-Neptune', all without clouds or hazes. For many of the Saturn and Jupiter-analog cases, refracted light could be detected in $<$10 hours of JWST time. This integration time can be achieved in 1 transit for Jupiter and Saturn-analog planets with T$_{eq}<$600 K orbiting F, G and K stars and in $<$5 transits for T$_{eq}<$400 K orbiting M dwarfs. For cases in which multiple transits are required, the total time from first transit to last is $<$1 year, and typically $<$6 months. Figure \ref{fig:inttime}A shows our results for $b$=0.0, with observing times required increasing by 1\% for $b$=0.2, 10\% for of $b$=0.6, and 30\% for $b$=0.9. 

%shows our results for an \textbf{impact parameter} ($b$) of 0.0, our results remain largely unchanged with impact parameter. The observing times required increase by 1\% for $b$=0.2, 10\% for of $b$=0.6, and 30\% for $b$=0.9.    .....changed this to get rid of the "largely unchanged" words since you don't need them any more now that you have quantified the change with b. 

%old version of paragraph
%\textbf{Figure \ref{fig:numtransits} shows the number of transits needed to detect refracted light for the cases shown in Figure \ref{fig:inttime}. Refracted light can be detect within 1 transit for planets with temperatures $<$500 K for Jupiter and Saturn analogs (with JWST) and N$_2$ Super-Earths (with E-ELT). Cooler ($<$300 K), cloudy planets with either H$_2$ or N$_2$-dominated atmospheres could also have detectable refracted light signals with 1 transit with E-ELT. Planets orbiting M dwarfs typically require 1-5 transits to detect refraction light because of the shorter transit duration. However, the transit periods are also shorter, and we have found that refracted light could be detected  in $<$6 months between first and last observations for most Jupiter and Saturn analogs (with JWST) and N$_2$ Super-Earths (with E-ELT).}

%The results shown assume that observations are made at 1 $\mu$m at one wavelength with R=100 for a planet at a distance of 10 pc.   ...what does "at one wavelength" mean here?  I removed it from the above.   

\begin{figure}[h]
\includegraphics[width=8cm]{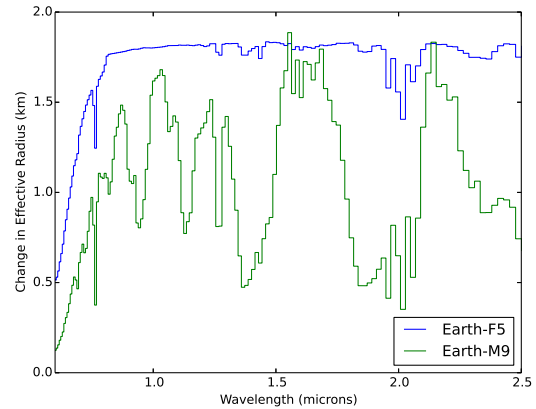}
\caption{Wavelength-dependent changes in the out-of-transit light curve for a 2 R$_{\oplus}$ Super-Earth with an Earth-like atmosphere, represented as the change in effective radius (km). The spectra shown are the differences in the spectra between Stage 2 and Stage 1 of the transit (see Figure \ref{fig:lightcurve}). This figure shows that the refracted light signal could be detected over a wide wavelength range, and that it should be possible to bin over multiple spectral resolution elements to reduce the integration time needed to detect refracted light.}
\label{fig:lightcurve-diff}
\end{figure}

Figure \ref{fig:inttime}B shows the E-ELT integration time required to detect refracted light for super-Earth and mini-Neptune atmospheres with $b$=0.0. We calculated the signal levels for N$_2$, H$_2$O, CO$_2$ and H$_2$ atmospheres, but only a comparison of N$_2$ and H$_2$ atmospheres is shown here. We find that refracted light could be detectable in $<$10 hrs of E-ELT time for many of the clearsky atmospheres, and even some cloudy atmospheres. In contrast, detecting refracted light for a hazy exoplanet would require $>$100 hours for all the planetary atmospheres we considered. Here we have assumed that it is possible to bin over at least 50 spectral resolution elements. The justification for this is found in Figure \ref{fig:lightcurve-diff}, which shows the wavelength-dependent refracted light signal for 2 R$_{\oplus}$ planet with an Earth-like atmosphere. A larger change in effective radius at a given wavelength means a stronger flux from refraction prior to ingress or after egress. Between 0.8 and 1.35 $\mu$m - shortward of a major H$_2$O absorption feature and where Rayleigh scattering opacities are small - there are $\sim$50 spectral resolution elements that could be summed. For Earth-analog atmospheres, there is a relatively large flux difference at all these wavelengths. Therefore, we consider binning over multiple spectral resolution elements to decrease the integration time to be a valid approach, at least for N$_2$-dominated planets like Earth.

% large amount of ->  large

%what do you mean by "wavelength units" here, spectral resolution elements?  

The greatest amplitude of refracted flux for the N$_2$ Super-Earth cases around a Sun-like star is 0.12 ppm for a 400 K planet. For the cloudy case, the maximum amplitude is 0.06 ppm at 200 K. The cloudy H$_2$ cases have amplitudes between 0.2 and 1.0 ppm, but only for the very cold ($<$200 K) cases. The amplitudes for the hazy H$_2$ cases are all below 0.025 ppm, and below 0.005 ppm for T$_{eq}>$150 K. The number of transits required to detect refracted light with E-ELT is 1 for clearsky N$_2$ Super-Earths with $T_{eq}<$800 K orbiting F, G and K stars and $<$3 for T$_{eq}<$500 K for those orbiting M dwarfs. As with the Jupiter and Saturn analogs, 3 transits is, from first transit to last, much less than a year and typically $<$3 months. Cloudy atmospheres with T$_{eq}<$250 K could exhibit detectable refracted light signals, but hazy atmospheres have largely undetectable refracted light signals except for some very cold (T$_{eq}$=100 K) cases.

%\begin{figure}[h]
%\includegraphics[width=8cm]{numtransits-all}
%\caption{\textbf{Number of transits needed to detect refracted light for planets with H$_2$-dominated atmospheres (with JWST) and Super-Earth/Mini-Neptune planets (with E-ELT). For planets orbiting F, G and K stars, refracted light can be detected within 1 transit for much of parameter space, and for planets orbiting M dwarfs, refracted light can be detected in $<$5 transits, which is typically $<$6 months. Here, as in Figure \ref{fig:inttime}, the Super-Earth/Mini-Neptune calculations have been performed under the assumption that 50 wavelength units can be binned over at R=100.}}
%\label{fig:numtransits}
%\end{figure}

\section{Discussion} 

A detection of refracted light implies a haze-free atmosphere because refracted light is much more detectable for a clearsky atmosphere than for a hazy one (see Figure \ref{fig:inttime}B). However, discriminating between cloudy and hazy worlds could be more challenging. For example, for a 600 K N$_2$ Super-Earth orbiting a Sun-like star (and for the majority of parameter space), a null detection of refracted light would be consistent with either a cloudy or hazy atmosphere, with no apparent way to differentiate between the two. On the other hand, for a 250 K N$_2$ Super-Earth orbiting a Sun-like star, both the clearsky and cloudy cases are consistent with a detection of refracted light. To disambiguate these results, one would need to quantify the refracted light, which would require more observing time. Overall, a detection of refracted light is indicative of a non-hazy atmosphere and, for some regions of parameter space, quantifying the refracted light flux could aid in uniquely discriminating between cloudy, hazy and clearsky atmospheres.

%one would need to not only detect refracted light, but also quantify its exact amount,  ->  one would need to quantify the refracted light.    
%since if you need to quantify it, then you have to have detected it first. 

The refracted light brightness is strongest for planets with T$_{eq}$ between 150-350 K, and is undetectable for very high temperature planets. For hot, close-in planets, the planet-star distance ($d$) is small, meaning that the characteristic deflection angle R$_*$/d is large, and that large refraction angles are required to produce a strong refracted light signal. For T$_{eq}>$800 K, angles this large would require probing pressures greater than 1 bar, where most atmospheres should be opaque, meaning that atmospheric opacity results in low refracted light signals. For the coldest (T$_{eq}<$150 K) planets, $d$ is large and the refraction angles for clearsky atmospheres are often much larger than R$_*$/d. This results in more refracted light being observed further away from ingress and egress, increasing the average flux in Stage 1 relative to Stage 2 and reducing the overall detectability. (see Figure \ref{fig:lightcurve}).

% Beautiful!! 

%Detecting refracted light should be possible in the near future with JWST and E-ELT. The best prospects for testing this method would be Saturn-analogs with temperatures between 150-350 K. \textbf{For these cases, refracted light could be detected in one transit with JWST for most stellar types.} Planets with \textbf{hot} equilibrium temperatures, such as the close-in Jupiter to Neptune-sized planets being characterized today, will tend to be poorer candidates for detecting refracted light because, as shown in Figure \ref{fig:inttime}, the refracted light signal is very difficult to detect for planetary equilibrium temperatures $>$800 K for planets orbiting Sun-like stars and $>$400 K for planets orbiting M dwarfs. \textbf{The reason for this is that the refracted light signal is detectable when the angle of refraction at the pressure cutoff is within a factor of a few of the angular size of the star. For very close-in planets, the angular size of the star is larger than the refraction angle at 1 bar, so the signal is not as detectable as for further out planets. }

In the near future, E-ELT could be used to identify non-hazy potentially habitable planets, which have 180$<$T$_{eq}<$260 K (\citet{kopparapu13}, Ravi Kopparapu, private communication). As shown in Table \ref{tab:results} and Figure \ref{fig:inttime}, refracted light could be detectable with one transit with E-ELT, or $<$5 hrs of out-of-transit E-ELT time for potentially habitable N$_2$ dominated Super-Earths orbiting F, G and K stars. For planets orbiting M dwarfs, the required number of transits is typically less than two, with a total integration time of $<$5 hrs. \citet{hedelt13} estimate that it could take up to 10 transits to detect H$_2$O and CO$_2$ for Earth-like planets orbiting F, G and K stars with E-ELT using filter photometry, and \citet{rodler14} find that it would take $>$20 hours of E-ELT time to detect O$_2$ for Earth-like planets orbiting M dwarfs using high resolution (R$>$10000) spectroscopy. These estimates are larger than the amount of out-of-transit E-ELT time necessary to detect refracted light for potentially habitable planets. Therefore, because refracted light could be more detectable than spectral absorption features, looking for refracted light to distinguish between hazy and non-hazy exoplanets could be a useful tool in selecting exoplanets for extended follow-up observations. 

\section{Conclusions}

Increases in out-of-transit flux due to refraction prior to ingress and subsequent to egress could be detectable with $<$10 hours of out-of-transit observing time for Saturn and Jupiter-sized planets with JWST and for Super-Earths/Mini-Neptunes with E-ELT. Detecting refracted light would be indicative of a haze-free atmosphere, and a quantification of the amount of refracted light could aid in distinguishing between cloudy and clearsky atmospheres for planets with equilibrium temperatures $<$300 K. Because refracted light can, in some cases, be detectable with less than a few hours of out-of-transit observing time, this method could be an economical way of determining if an exoplanet is haze-free and therefore a good target for extended follow-up observations.

\section*{Acknowledgments}

This work was performed by the NASA Astrobiology Institute's Virtual Planetary Laboratory, supported by the NASA Astrobiology Institute under Cooperative Agreement solicitation NNH05ZDA001C.

\end{document}